\documentstyle[aps,preprint]{revtex}
\begin{document}

\draft
\title{Phenomenological Theory of Survival}

\author{Mark Ya. Azbel'}

\address{School of Physics and Astronomy,Tel Aviv University,\\
         Ramat Aviv, 69978 Tel Aviv, Israel}
\date{\today}
\maketitle

\begin{abstract}
Theoretical analysis proves that human survivability is dominated by an
unusual physical, rather than biological, mechanism, which yields an exact law. 
The law agrees with all experimental data, but, contrary to existing theories, 
it is the same for an entire species, i.e., it is independent of the population,
its phenotypes, environment and history.  The law implies that the survivability 
changes with environment via phase transitions, which are simultaneous for all 
generations. They
allow for a rapid (within few percent of the life span) and significant
increase in the life expectancy even above its value at a much earlier age.

\end{abstract}
\newpage
\narrowtext
Mortality is one of the most universal and important phenomena in biology.
Human mortality is extensively studied over two centuries \cite{1,2,3,4,5}  and is arguably
the  best statistically quantified  biological phenomenon.  There exist several 
evolutionary theories of aging \cite{6} (the first one \cite{7} is only 50 years old). 
Yet, C. Franceschi (in ref.\cite{8}) notes: "Longevity is a trait with some 
peculiarities because of the unnecessary nature of aging".  This paper 
proves that, in contrast to all existing theories, human mortality is dominated 
by an unusual physical (rather than biological) mechanism.  Presumably, the same 
mechanism dominates mortality of laboratory fly populations.

 I study a large
and diverse amount of over 150,000 data points from 1528 period life tables,
for 3 races on 4 continents in 16 countries during over a century of their
history, 36 cases total (here and on ''a case'' denotes the population of a
given sex in a given country at any time in its history; in the USA white,
black and total populations are considered as separate cases). Each period
life table presents in any given case the probability $l_x$ to survive to any
(yearly) age x in a given calendar year. To escape any arbitrary scaling or
adjustment, I use life table variables only.

The survivability $l_{x}$ depends on age $x$, calendar year $t$ and a complete
set $A$ of all parameters, comprehensively describing a considered population,
i.e. its phenotypes and their heterogeneous social, medical, dietary, etc. conditions during the
previous $x$ years (from $t$ - $x$ till $t$). So,$l_{x}=l_{x}(t,A)$. Mortality rate,
and thus its statistics and accuracy, are low in early age. The number of
survivors, and thus accuracy, decrease in old age. So, middle age
survivability, e.g., $l _{40}$, is determined more accurately. Present 
$l_{x}(t,A)$ and $l_{40}(t,A)$ from the same life table as the ordinate and
abscissa of a point. According to Fig. 1, for different races, countries, continents,
histories  $l_x$  is dominated 
by  the function $l_x(l_{40})$, which depends only on  $l_{40}$, i.e. invariant to $A$ and 
is therefore the same for an entire species.  The deviation of  $l_x$  from this 
function depends on $A$ (and thus on population phenotypes, living conditions, 
history).  It is relatively small. From now on I denote  $l_x(l_{40})$ as "the invariant
survivability" and, unless specifically stated otherwise, consider this (dominant)
fraction of survivability only. 

By Fig. 1, at any age and in every case, $l _{40}$
 significantly changes with living conditions. Thus, since living conditions
of different groups in a given case are different, these groups may have
different probabilities $l_{x}^{\prime }$ to survive to $x$ years (the value
of $l_{x}$ in the population is the average $l_{x}^{\prime }$; $0\leq
l_{x},l_{x}^{\prime }\leq 1$). Suppose the probability density of a given $%
l_{40}^{\prime }$\ in the population is c($l _{40}$,$l_{40}^{\prime }$%
), i.e.

\begin{equation}
\int\limits_{0}^{1}c(l_{40},l_{40}^{\prime })dl_{40}^{\prime }=1,\qquad
l_{x}(l_{40})=\int\limits_{0}^{1}l_{x}^{\prime }\cdot
c(l_{40},l_{40}^{\prime })dl_{40}^{\prime }  \label{1}
\end{equation}

By Eq. \ref{1}, $0\leq \min (l_{40}^{\prime })\leq l_{40}\leq \max
l_{40}^{\prime }$.So, $l_{40}^{\prime }=0$ when $l_{40}^{{}}$ = 0 and $%
l_{40}^{\prime }=1$ when $l_{40}=1$. Thus, invariance, which allows one to
introduce c and which yields Eq. \ref{1}, implies that any population is
homogeneous in $l_{40}^{\prime }$ at the distribution boundaries. In virtue
of invariance, the dependence of $l_{x}^{\prime }$\ on $l_{40}^{\prime }$\
is the same as the dependence $l_{x}(l_{40})$, i.e., $l_{x}^{\prime
}=l_{x}(l_{40}^{\prime }).$ This, by Eq. \ref{1}, yields a remarkable
symmetry of the invariant survivability to the transformations, specified by
the function $c$ of two variables:

\begin{equation}
l_{x}\left[ \int\limits_{0}^{1}l_{40}^{\prime }\cdot c(l_{40},l_{40}^{\prime
})dl_{40}^{\prime }\right] =\int\limits_{0}^{1}l_{x}(l_{40}^{\prime })\cdot
c(l_{40},l_{40}^{\prime })dl_{40}^{\prime }  \label{2}
\end{equation}
Such invariance is sufficiently restrictive to mathematically accurately 
determine  $l_x(l_{40})$ (as an invariant of survivability dynamics with respect 
to any population and environmental changes). Indeed,
in a general case the function $l_{x}(l_{40})$, which yields Eq. \ref{2},
must be linear. A special case of $c(l_{40},l_{40}^{\prime })=c^{\prime }\cdot \delta
(l_{40}^{\prime }-l^{\prime })+c^{"}\cdot \delta (l_{40}^{\prime }-l^{"})$%
yields $c^{\prime }+c^{"}=1$ by Eq.\ref{1}, and $l_{x}(l_{40})=l_{x}(c^{%
\prime }\cdot l^{\prime }+c^{"}\cdot l^{"})=c^{\prime \cdot }l_{x}(l^{\prime
})+c^{"}\cdot l_{x}(l^{"})$ by Eq. \ref{2}. The latter equation implies 
$dl_{x}/dl_{40}=dl_{x}/dl_{x}^{\prime }=$const. On the other hand, a linear 
$l_{x}(l_{40})$ yields Eq. \ref{2} with any arbitrary $c$ in virtue of Eq. \ref{1}.
Similarly, if an entire population is distributed in a certain interval of $%
l_{40}$, then $l_{x}(l_{40})$ is linear in this interval, and the population
is homogeneous at its ends (as it was in the previous case).
Correspondingly, if any population is homogeneous at certain $l_{40}$
points only, then an entire population is distributed within one of the
intervals $\lambda _{s}<l_{40}<\lambda _{s+1}$ (s = 0,1 is its ordinal
number): 
\begin{equation}
\bigskip \int\limits_{\lambda _{s}}^{\lambda _{s+1}}c(l_{40},l_{40}^{\prime
})dl_{40}^{\prime }=1;\qquad l_{40}=\int\limits_{\lambda _{s}}^{\lambda
_{s+1}}l_{40}^{\prime }\cdot c(l_{40},l_{40}^{\prime })dl_{40}^{\prime }
\label{3}
\end{equation}
and $l_{x}(l_{40})$ is linear within these intervals. When $l_{40}=\lambda
_{s+1}$, an entire population of all ages moves to the next interval.
Clearly, such piecewise $l_{x}(l_{40})$ satisfies Eq.\ref{2}. When $\lambda
_{0}=0,\lambda _{1}=1,$ this case reduces to the previous one. If an
infinite number of edge points in Eq.\ref{3} condenses and forms a
continuous interval (where any population is homogeneous), then the
invariant $l_{x}(l_{40})$ in this interval remains undetermined. 
Of course, the possibility to distinguish this case from the case of 
a large number of short linear segments is limited by the disregarded 
non-invariant contributions to survivability, but demographic studies
prove that survivability is heterogeneous and exclude the former case 
(of a population which is homogeneous in a finite interval). Thus, in a
general case 
\begin{equation}
l_{x}=R_{x}^{(s)}l_{40}+a_{x}^{(s)},\qquad \lambda _{s}\leq l_{40}\leq
\lambda _{s+1}.
\end{equation}
So, except for the number of linear segments,  the
very existence of the invariant survivability allows one to establish its exact
law, without any experiments, approximations, and assumptions.\\ 
Equation (4) is the implication of invariance only. Such invariance was also
demonstrated \cite{9} for  medfly and fruitfly
families whose different populations were extensively studied in different
conditions. Fly statistics is rather low, thus their $l_{x}$ was studied as
a function of the life expectancy at birth (which averages $l _{x}$
over different ages and is therefore more smooth). The study was based on
fly populations hatched the same day. However, since in most cases fruitflies
were kept in stationary conditions, it suggests that the invariant
survivability law is general for humans and flies.

Piecewise linear law (4) and its age independent invariant intersections are
the main predictions of this paper. They are verified with all experimental
data - see, e.g., $l_{1}$, $l_{60}$,$l_{80}$ vs $l_{40}$ in Fig.2.
Piecewise linear law agrees with (but has never been suggested in)
demographic approximations \cite{3}. Slope jumps in Eq.(4) and Fig.2 are consistent with
ref \cite{5}. However, in contrast to the  qualitative observations in ref.\cite{5}, accurate
Eq.(4) yield quantitative predictions. Since all $l_{x}$ belong to the same
calendar year, their slope jumps are simultaneous for all generations.
Indeed, although the intersections in Fig.2 are reached at different times
in different countries for different sexes, they are simultaneous (to the
invariance accuracy) at different ages, and (in $l_{40}$) for different populations. 
The invariant law allows for the
prediction of intersections. For instance, the extrapolation of $l_{80}$
beyond $\lambda _{1}=0.95$ in Fig. 1 yields $l_{40}=1$ (while $l _{80}$
is finite), i.e. implies no deceased until $40y$. This is hardly possible, and
suggests a crossover, which is indeed seen in Fig.2 at $\lambda _{2}=0.97$
The extrapolations of $l_{60}$ and $ l_{80}$ to $l_{40}<0.94$ yield 
$l_{60}=0$ and $l_{80}=0$\ , i.e., no survivors beyond $60y$ (when $l_{40}=0.27$)
 and $80y$ (when $l_{40}=0.42$). This is never true, and suggests an
intersection at $0.4<l_{40}<0.7$.

To elucidate the nature of the invariant law, present Eq. (4) in a different
form:

\begin{equation}
\bigskip l_{x}=c_{x}l_{x}^{(s)}+(1-c_{s})l_{x}^{(s+1)}~\quad when~\quad
l_{x}^{(s)}\leq l_{x}\leq l_{x}^{(s+1)}
\end{equation}

Here $l_{40}^{(s)}=\lambda _{s};l_{x}^{(s)}=l_{x}(\lambda _{s})$ \ from
Eq.(4); $c_s$ reduces to $l_{x}^{(s)}$\ and $l_{40}$ [thus, by Eq. (3), to c].
Equation (5) accurately separates ''nature'' and ''nurture'' in survival.
''Nature'' reduces to the fixed set of the intersection survivabilities .
The set depends only on age and is invariant, i.e., independent of
phenotypes (and thus of a specific DNA sequence at least in an entire
species), their living conditions and life history. The dependence of the
set on age is not determined by invariance, but invariance implies that it
is the same for at least an entire species. (Moreover, it scales onto the
same functions for species as remote as humans and flies \cite{9}). Thus, it
does not change at least as long as the species does not evolve into a
different species (demographic data in Fig.2 verify it for 100-150 years).
So, the set must be inheritable. Such set reminds of the body temperature
(which in any living conditions is the same with few percent accuracy) of an
entire class of birds and a subclass of placental mammals. Presumably, both
the set and the body temperature are genetically determined, and independent
of a specific DNA sequence. But the set, unlike the body temperature,
strongly depends on age. ''Nurture'' distributes the survivability $l_{x}$
at a given age between two adjacent intersection survivabilities. The
concentration $c_{s}(0\leq c_{s}\leq 1)$ is age independent (and may be
related to, e.g., $l_{40}$). Thus, survivability follows environment (in
particular, a new intersection survivability emerges) simultaneously for all
generations.

Equation (5) relates $c_{s}$, and thus $l_{x},$ to $l_{1}$. Since  $l_{x}$ is the survival
probability in the same calendar year as $l_{1}$, so, by Eq. (5), the
survivability accurately and rapidly  follows the change in
environment according to the value of $l_{1}$ (i.e., the infant mortality $%
q_{0}=1-l_{1}$), which is established in less than two years. Indeed, whatever the difference in
environmental factors is, close values of infant mortality imply very close
fractions of deceased at any given age in 1885 Swedish and 1947 Japanese
females, despite of their different races, continents, countries, and 62
year gap in their different history. Since $l_{1}$ depends on the
environment in a given year only, the invariant survivability $l_{x}$ is
reversible (an entire survival curve comes back when $q_{0}$ changes
non-monotonically and returns to its previous value), and statistically
independent of the life history during its $x$ years, despite country
specific, highly and irregularly changing, non-monotonic and non-stationary
living conditions. World wars and epidemics, e.g., flu in 1918 Europe,
significantly decrease $l_{x}$. (For instance, in 1915 the probability for a
French male to survive to $80y$ was 5 times less than in 1913, twice less than
in 1917, and 3.5 times less than for a 1861 Swedish male). Yet, they just
slightly shift the plots (mostly vertically, and relatively little) in
Fig.2. In a couple of years (which estimate the relaxation time at few
percents of the life span) all plots restore their invariant dependence,
i.e., the memory of the previous life history is erased. Accurate
reversibility of the invariant survivability does not decrease with aging,
even in old age. Unless such reversibility is related to some perfect
biological rehabilitation (which is hardly possible), it implies an
adiabatic change in a certain thermodynamic equilibrium. This is consistent
with its relaxation time (rapid compared to the life span, but enormous on a
microscopic scale). Equation (5) reduces $l_{x}$ to the fixed set of $%
l_{x}^{(s)}$. Since $l_{x}^{(s)}$ reversibly change into each other, they
are related to different equilibrium thermodynamic states of the same
system, i.e. to different phases. Thus, Eq. (5) relates the invariant
survivability to a certain phase equilibrium, and the jumps in Fig.1 slopes
to the emergence of a new phase. The phase concentration $c_{s}$, which
Eq.(5) reduces to, e.g., $l_{1}$, is independent of age. This suggests that
the age dependence of a \ $l_{x}^{(s)}$ is related to the difference in
survivabilities provided by the same phase at different ages. Quasi-equilibrium 
phases in a living homeostatic being might be
related to a meso- (e.g., a cell) or microscopic (e.g., DNA
configuration) scale. 

While the existence of survivability phases is accurately proven, their
microscopic nature, as well as that of the parameter which determines their
concentrations, remains unknown (as ''units of heredity'' were to Mendel).
 However, any phase equilibrium may be reversibly
manipulated. This means that human life expectancy may be rapidly (within
few years) reversed to its value at a much earlier age. (Note that in just 8
years from 1947 till 1955 the life expectancy of Japanese females increased
26\% at birth, 15\% at $60y$, and 20\% at $80y$).

Presumably, the change in phase concentrations may affect other than
survivability characteristics, in particular, aging and disease resistance.
Indeed, Eq.(5) is not violated even in Japan prior to 1949 (see Fig.1) and
in Finland from 1890 till 1940 (see Fig.2), although their mortality has a
strongly tubercular age pattern during this period.

The suggested phase equilibrium nature of survivability implies that
non-equilibrium (in particular, sufficiently non-stationary and
heterogeneous conditions which depend on age, sex, social mobility,
immigration and other factors), may lead to more than two adjacent phases
and to hysteresis in the adjustment. The contribution of non-equilibrium
phases is most pronounced in old age, when the difference in $l_{x}^{(s)}$
is significantly higher. This agrees with Fig.2. Hysteresis yields small
jumps in $l_{x}$ and shifts of the intersections, which are indeed present
in certain cases. The accuracy of the invariant law estimates maximal
concentrations of ''extra'' phases. They are lower for females, presumably
suggesting lower heterogeneity of female populations.

Survivability independence of life history implies no correlation between
early and old age invariant mortalities in a given cohort (born the same
year). This disagrees with evolutionary theories of aging \cite{6}.
Stochastic mutation accumulation \cite{7} theory is inconsistent with reversibility
and rapid accurate survivability change with environment. Optimal allocation 
of metabolic resources (pleiotropy and
disposable soma theories) implies  strong
correlation between survival in young and old age in populations born the same year. This
is inconsistent with the survivability independence of life history.
Reversibility is also inconsistent with mortality theories (thelomers,
oxygen consumption, free radicals, somatic mutations). Although natural
mortality in the wild is mostly due to extrinsic hazards, invariant
mortality, which dominates in species as remote as humans and flies (in
certain protected environment), and which rapidly, accurately and
simultaneously in all generations changes with environment, calls for
biological, evolutionary and microscopic physical theories.\\
These conclusions, as well as an unusual mechanism of mortality, are accurate
implications of the exact law, which agrees with all demographic approximations 
and studies \cite{3,5}.  For instance, the infant mortality is widely appreciated by 
demographers as a sensitive barometer of environmental conditions. However,
demographic approximations are developed primarily as a useful tool of maximally
accurate estimation and forecast.  They are often country and time specific, and 
approximate specific demographic data better than Eq. (4), but they do not consider 
nor care about the underlying general law and its mechanism.  In contrast, my goal 
is the exact law, albeit of the invariant mortality only (to discover inertia, one
must disregard friction!) and its mechanism, which yield new biological insights.

\begin{description}
\item[Acknowledgement]  
\end{description}

\vspace{0in}This paper was stimulated by enlightening discussions with
Profs. Y. Aharonov, S.M. Jazwinski, S. Horiuchi,  D.S. Thaler, A. Libchaber,and 
B. Tsirelson. I am grateful to Prof. S.
Horiuchi, S. Smallwood, A. Hanika, K. Anderson, S. Helgadottir, C. Capocci,
E. Waeyetens, N. White, U. Wegmueller, V. Kannisto, M. Nieminen, K. Deaves
for life tables; to I. Kolodnaya for crucial technical
assistance; A. Kobi for computer simulations, J. \& R. Meyerhoff chair for
financial support.

\begin{figure}
\caption {Probabilities (vertical axis, upper, middle and lower curves correspondingly)
to survive to 1, 60, 80 years vs $l_{40}$ (horizontal axis) for females in 1909-1997 
Australia, 1880-1998 Belgium, 1891-1996  Japan, 1861-1995  Sweden.
Swedish and Japanese females are denoted by full square, open square, and 
full triangle, open triangle; others by full circle, open circle; open signs denote 
1914-1919; 1939-1947. Solid lines are linear regressions, which verify predicted 
piecewise linearity. }
\label{f1}
\end{figure}

\begin{figure}
\caption {Same as in Fig. 1, but for all cases (males and females in 1909-1997
Australia, 1880-1998 Austria; 1880-1998 Belgium, 1950-1987 Canada and provinces;
1851-1998 England and Whales; 1881-1998 Finland, 1898-1995 France; 1871-1994 Germany;
1841-1998 Iceland; 1925-1992 Ireland; 1891-1996 Japan, 1846-1998 Norway; 1930-1992 Scotland;
1861-1995 Sweden, 1878-1993 Switzerland; 1900-1997  USA, white, black, and total population.) 
To amplify invariance and piecewise linear dependence, some of the linear segments are slightly 
rotated and shifted. (This does not violate piecewise linearity.)} 
\label{f2}
\end{figure}

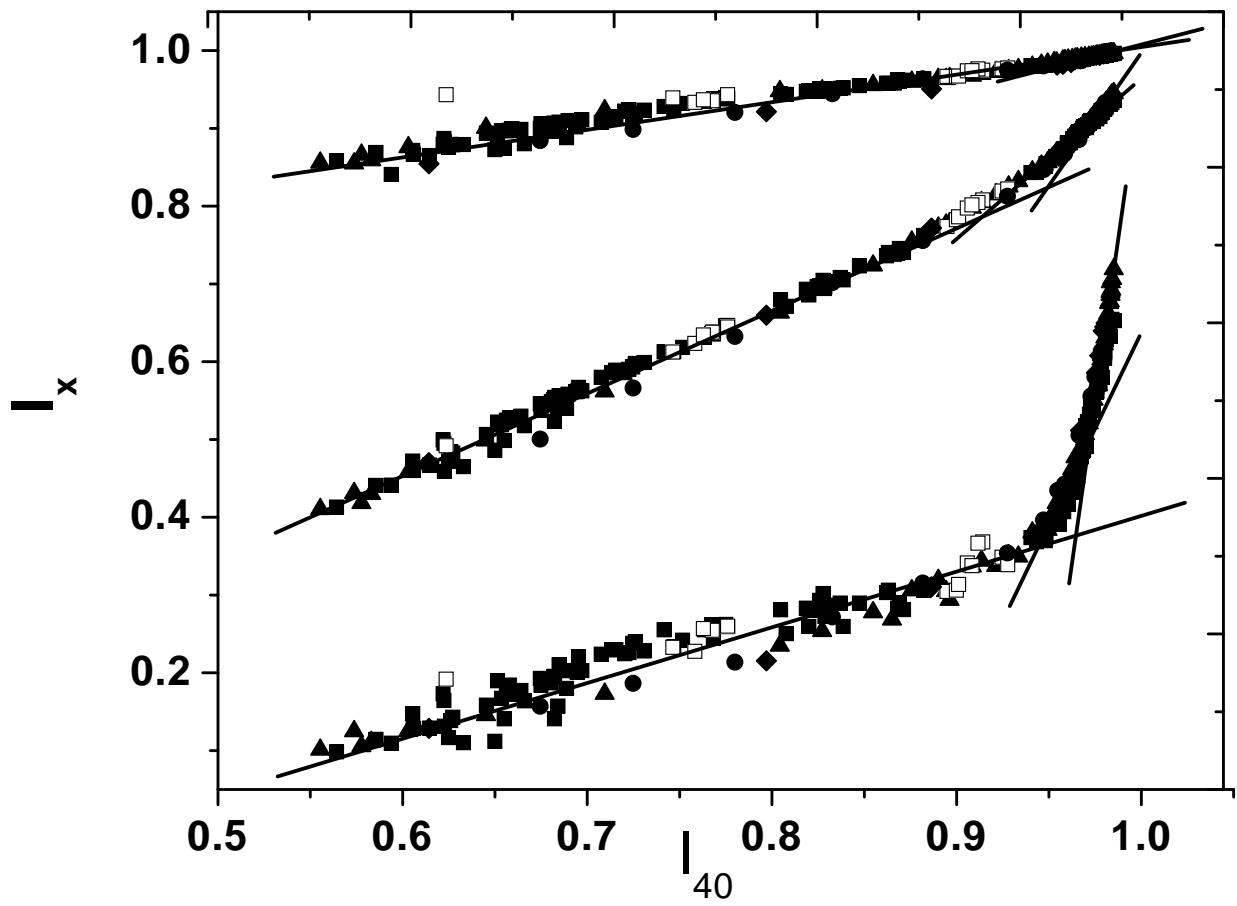
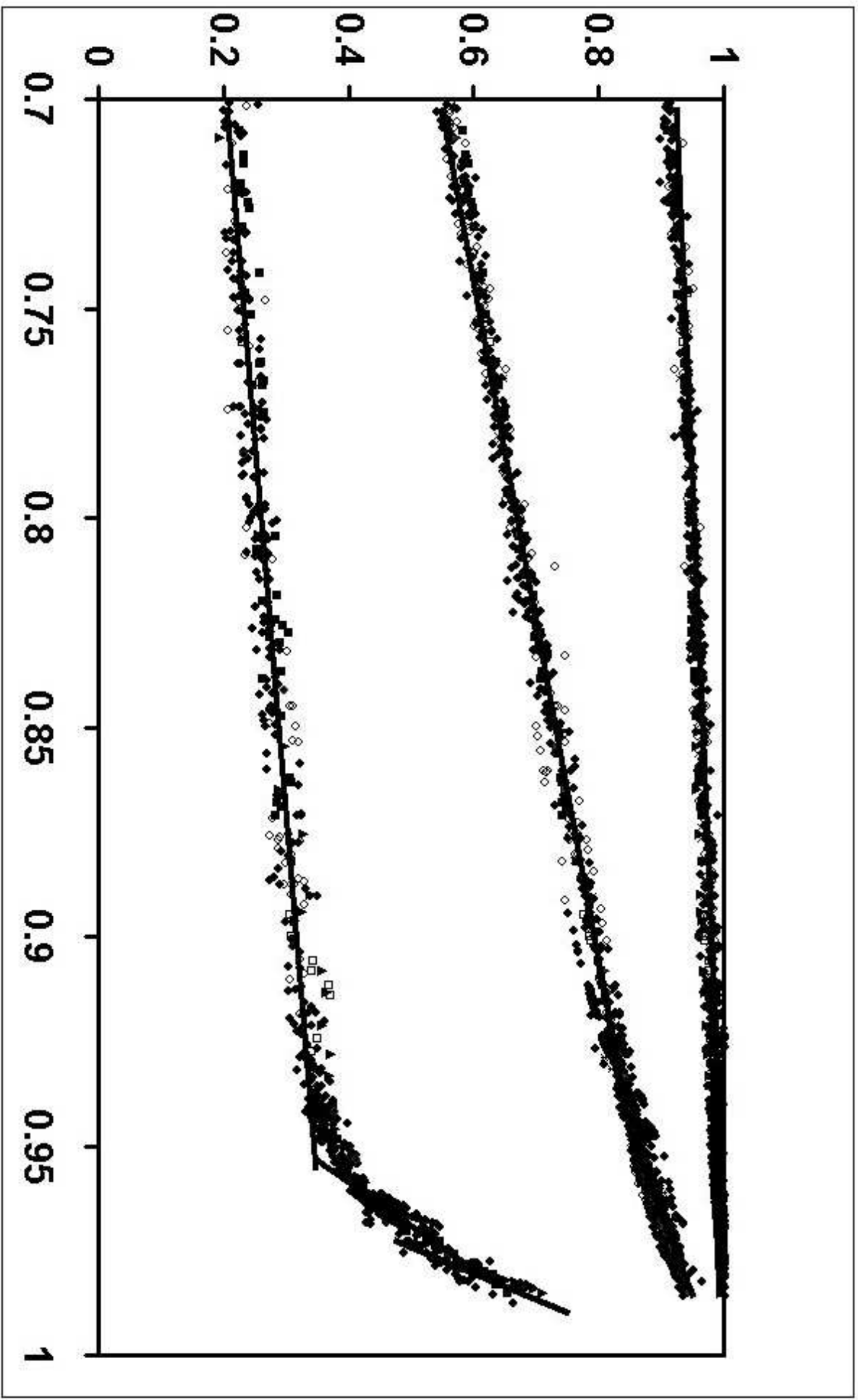
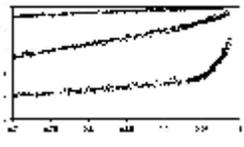
\end{document}